\documentclass{article}
\usepackage[english]{babel}
\usepackage{theorem}
\usepackage{makeidx}
\usepackage{amsmath}
\usepackage{epic}
\usepackage{amscd}
\usepackage{curvesls}
\usepackage{psfig}
\usepackage{bbm}
\usepackage{brakets}
\usepackage{amssymb}
\usepackage{rotating}
\newtheorem{defn}{Definition}[section]

\newtheorem{nota-rem}[defn]{Notation--Remark}

\renewcommand{\thefootnote}{\dag}

\begin{document}
\title{A nonperturbative\\ Real-Space Renormalization Group scheme}
\author{Andreas Degenhard\thanks{e-mail: hanne@physik.uni-bielefeld.de}\\
         Department of Mathematical Physics, University of Bielefeld,\\
         Universit\"atsstra\ss e 25, D-33615 Bielefeld, Germany\\[0.6cm]}
\date{}
%
%
\maketitle
\begin{abstract}
  Based on the original idea of the density matrix renormalization group (DMRG)
  \cite{wh93}, i.e. to include the missing boundary conditions between adjacent
  blocks of the
  blocked quantum system, we present a rigorous and nonperturbative mathematical
  formulation for the real-space renormalization group (RG) idea invented
  by L.P. Kadanoff \cite{ka66} and further developed by K.G. Wilson \cite{wi71}.
  This is achieved by using additional Hilbert spaces called auxiliary spaces
  in the construction of each single isolated block, which is then named a
  superblock according to the original nomenclature \cite{wh93}. On this superblock
  we define two maps called embedding and truncation for successively integrating out
  the small scale structure. Our method overcomes the known difficulties of the
  numerical DMRG, i.e. limitation to zero temperature and one space dimension.
\end{abstract}
\vskip 8mm
%
\section{Introduction}
Soon after K.G. Wilson's dramatic success in applying a
momentum space formulation of the renormalization group (RG) method
\cite{ka66} to the Theory of Critical Phenomena and the Kondo Problem
\cite{wi75} there was a considerable amount of efforts in applying the same
type of approach as a real-space formulation to a variety of quantum physical problems.
Since the momentum space formulation, apart from a few exceptions
\cite{{wi71},{wi75}}, relies in most cases on a perturbative expansion, real-space
methods offer non perturbative approaches and are therefore extremely important in
applying RG ideas to strongly and complex correlated systems.
It then turned out that for a variety of such physical models the real-space
RG techniques give considerable bad results and the reason was unknown for nearly
fifteen years. During that time some new real-space RG methods were discovered
and some of them worked out very well whereas other methods failed without giving
any insight to their failure. We like to refer the interested reader to the
book of T.W. Burkhardt and J.M.J. van Leeuwen \cite{bl82} for a summary of work
on this topic.\\
Apart from these developments S.R. White and R.M. Noack published a series of papers
containing a new idea for improving real-space RG techniques \cite{{wh93},{wn92}}.
Based on the understanding of the importance of boundary conditions for isolated blocks
in real-space RG methods for quantum physical systems a numerical approach was
invented to take sufficiently
many boundary conditions into account during the RG procedure. Apart from
the impressive accuracy of the numerical results this new approach
displays also the typical universal character of a RG formulation, in that it
is applicable with some particular changes for a variety of problems and was
named the Density Matrix RG (DMRG) \cite{{wh93},{wn92}}.\\
The dramatic success of the DMRG method has changed the picture of real-space RG
techniques completely and has been applied until now in very different fields of
scientific research \cite{{bb98},{cm96},{hh97}}. The method itself is a rather
complicated algorithm and a detailed description together with some examples
is given by S.R. White \cite{wh93}.\\
Despite of all the excitement concerning DMRG, the method has some important
restrictions which are given by the method itself and therefore cannot be removed
by applying simple changes to the DMRG algorithm. Here we mention the three
main restrictions briefly:
\begin{enumerate}
\item The chief limitation of DMRG is dimensionality. Although higher
  dimensional variations are not forbidden in general, it becomes a complicated
  task. Recent applications of DMRG to finite width strips in two dimensions
  show a declining accuracy with the width. Therefore a successful approach for
  two dimensions in general or even higher dimension has never been worked out.
\item DMRG is by definition an algorithm and therefore it is a purely numerical
  RG approach. Although this needs not to be a disadvantage we like to have an
  analytical formulation of the DMRG method. In such a reformulation the
  numerical DMRG scheme will occur as one possible realization of a more
  general description. We therefore expect a deeper insight to successful
  working RG approaches.
\item DMRG is restricted to zero temperature and is usually applied for
  calculating ground state properties like the ground state magnetization or
  even the ground state itself. Finite temperature results were obtained only
  in the low lying spectrum but with very limited accuracy. In comparison
  to other real-space RG methods DMRG is different because it is designed to
  calculate ground state quantities. Recently, based on the idea of
  Xiang et al, a thermodynamic method was applied successfully, which
  combines White's DMRG idea\cite{{wh93}} with the
  {\it transfer-matrix} technique\cite{krs99} and which is now called TMRG.
  Although TMRG is also purely numerical since it shares the basic idea with
  DMRG it is an even more complicated algorithm \cite{krs99}. Due to the
  close relationship to DMRG, the aim of TMRG is to give numerical accurate
  results for physical quantities and does not predict a RG flow-behaviour.
  In contrast our method is suited to calculate the flow-behaviour of
  the system, even analytically, although the main advantage by comparison with
  TMRG is the simple structure of our RG scheme. This makes it an easy task to
  apply it to a great variety of physical models.
\end{enumerate}
The rest of this article is organized as follows: In the next section we review
the key idea of DMRG shortly. We begin by introducing the standard concepts of
the real-space RG method in the language of spin chains in the way it was
originally proposed.
In section III we present a rigorous formulation of a real-space RG transformation.
Each single block within the blocked chain is enlarged by an additional space,
the auxiliary space. A single block together with its auxiliary space is called
a superblock for which a real-space RG transformation is defined by integrating
out the small spatial structure. Constructing a global RG transformation for the
complete quantum system from concatenation of the local superblock RG
transformations leads to the definition of exact and perfect RG transformations.
In section V we give some final remarks including the relation to previous
approaches in this direction. Applications in terms of this new formulation are
shifted completely to a second paper.
%
\section{The idea of DMRG}
\label{dmrgintro}
\begin{figure}[h]
\vspace{0.6cm}
\centerline{
  \psfig{figure=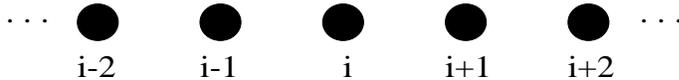,width=9.0cm,height=1.0cm}}
\vspace{0.8cm}
\caption{\em A one dimensional spin chain.}
  \label{1Dchain}
\end{figure}

\begin{figure}[h]
\vspace{0.6cm}
\centerline{
  \psfig{figure=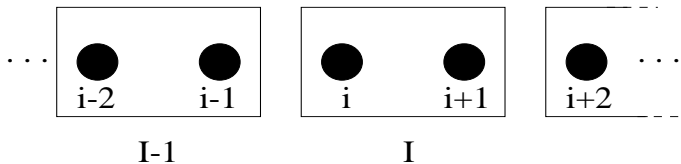,width=9.0cm,height=2.1cm}}
\vspace{0.8cm}
\caption{\em A one dimensional chain divided into blocks where each block is
  composed of two single sites.}
  \label{blocks}
\end{figure}

\begin{figure}[ht]
\vspace{0.6cm}
\centerline{
  \psfig{figure=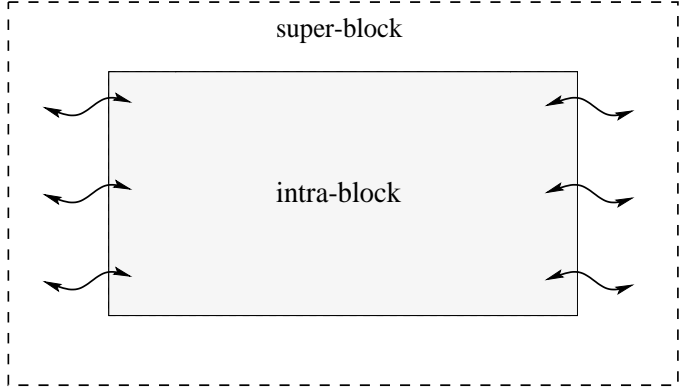,width=9.0cm,height=5.2cm}}
\vspace{0.8cm}
\caption{\em An isolated system block embedded into a superblock
             visualized by dashed lines.}
  \label{superblock}
\end{figure}

%
The very standard real-space RG approach is best explained for a
spin Hamiltonian $H$ on a one dimensional lattice as visualized in figure
\ref{1Dchain}.
%
%
The dots represent the individual spins which are grouped
together by breaking up the chain into blocks as visualized in figure
\ref{blocks} for a particular block composition of two sites.
%
%
We like to establish a notation in which small letters
refer to the single site spins and capital letters denote the blocks. The
block Hamiltonian for the block with the index $I$ is then denoted as
$H_I$. The idea of real-space RG is then to replace each block of the single
spins by one effective {\em block-spin}, which leads to a renormalized block-spin
Hamiltonian $H_{I'}$. The calculation of the block-spins from the blocks
composed of single site spins is carried out by a {\it RG transformation} $R$,
which can be defined in various ways \cite{bl82}, for example by projecting the
block on the low lying spectrum \cite{wh93}.
In summary a RG approach is designed to split of the whole system into
subsystems called blocks for which it is possible to reduce the degrees of
freedom. Iterating this procedure leads to a {\em RG flow} in the parameter
space of the model and the hope is to find a fixed point of this flow behaviour.
Such a fixed point Hamiltonian is helpful to determine the universal
behaviour of the physical model.\\
As explained in the introduction, the boundary conditions of a block within
the quantum system are essential for the calculation of a {\em RG step},
which is defined as one application of the RGT. In fact the different boundary
conditions represent the correlations in the quantum chain between
adjacent blocks to which we refer in the following as {\em system blocks}. To
provide the opportunity to choose those boundary conditions, which result in
the most accurate representation of an isolated block, the fundamental idea of
DMRG is to embed the system block into a ``bigger'' block, called {\em superblock}.
This nomenclature as well as the term system block is due to the original
work of S.R. White \cite{wh93}. In some sense this simulates the environment
represented by the surrounding spin sites and effectively smoothes out the sharp
effects of the boundary conditions, as depicted in figure \ref{superblock}.\\
%
%
To construct a working approach out of this overall picture we are immediately faced
with a twofold basic problem: How can we describe the embedding of the system
block within the superblock and how can one include the boundary conditions
during a RG step. In the framework of DMRG these problems are overcome by
focusing on one particular state, the {\it target state} $\Ket{\psi}$, which
is the ground state of the superblock Hamiltonian obtained by
diagonalization. By using a complete set of eigenstates of the system block
$\left\{\Ket{\psi^{\text{system}}_m},\; m=1,\dots ,{l_{\text{system}}}\right\}$
and a complete set of eigenstates for the environment represented by the superblock
$\left\{\Ket{\psi^{\text{environment}}_n},\; n=1,\dots ,{l_{\text{environment}}}\right\}$
we decompose the target state $\Ket{\psi}$ according to
\begin{equation}
\label{dmrg3}
 \Ket{\psi}\,=\,\sum_{m}^{l_{\text{system}}}\,\sum_{n}^{l_{\text{environment}}}
 \; c_{m,n}\,\Ket{\psi^{\text{system}}_m}\,
  \otimes\,\Ket{\psi^{\text{environment}}_n}\,.
\end{equation}
We are interested in those states, which lead to an optimal representation
of the target state $|\psi >$ in a ``truncated'' basis.
Of course in this way we loose the exactness of relation (\ref{dmrg3}) and we
therefore denote the new result as an {\em optimal approximation} expressed as
\begin{equation}
\label{dmrg4}
 \Ket{\psi}\,\approx\,\Ket{\psi^{\text{opt}}}\,=\,\sum_{p}^{l_{\text{opt}}}\,
    \sum_n^{l_{\text{environment}}}\; \gamma_{p,n}\,
     \Ket{\psi^{\text{opt}}_p}\,\otimes\,\Ket{\psi^{\text{environment}}_n}\,,
\end{equation}
where the {\em optimal states}
$\left\{\Ket{\psi^{\text{opt}}_p},\; p=1,\dots ,l_{\text{opt}}
  <{l_{\text{system}}}\right\}$ are defined in terms of the original
system block states by
\begin{equation}
\label{dmrg5}
 \Ket{\psi^{\text{opt}}_p}\,=\,\sum_{m}^{l_{\text{system}}}\;
  {\alpha}_{m,p}^{\text{opt}}\,\Ket{\psi^{\text{system}}_m}
\end{equation}
with some coefficients ${\alpha}_{m,p}^{\text{opt}}$.
The coefficients $\gamma_{p,n}$ in (\ref{dmrg4}) can be determined by examining
the {\em reduced density matrix} of the system block within the superblock
\cite{wh93}.\\
The twofold problem introduced above is therefore solved as follows: First the
embedding of the system block within the superblock is achieved by
reconstructing the target state of the superblock in a basis, in which the basis
vectors are given as a tensor product composition of states of the system block
and the chosen environment. In this way the system block is described within
the bigger superblock. Since we have not truncated the set of the states which
belong to the environment, the RG step for the system block is performed by
taking all possible boundary conditions within the selected environment into
account.\\
From the previous discussion it becomes obvious that the coefficients
$\gamma_{p,n}$ can only be determined numerically within real applications of
this technique. To develop a complete analytic approach, a method for using a target
state will be impractical and we can only use the overall picture represented in
figure \ref{superblock}.
%
%
\section{A rigorous real-space RG transformation}
\label{rigorous}
\begin{figure}[ht]
\vspace{0.6cm}
\centerline{
  \psfig{figure=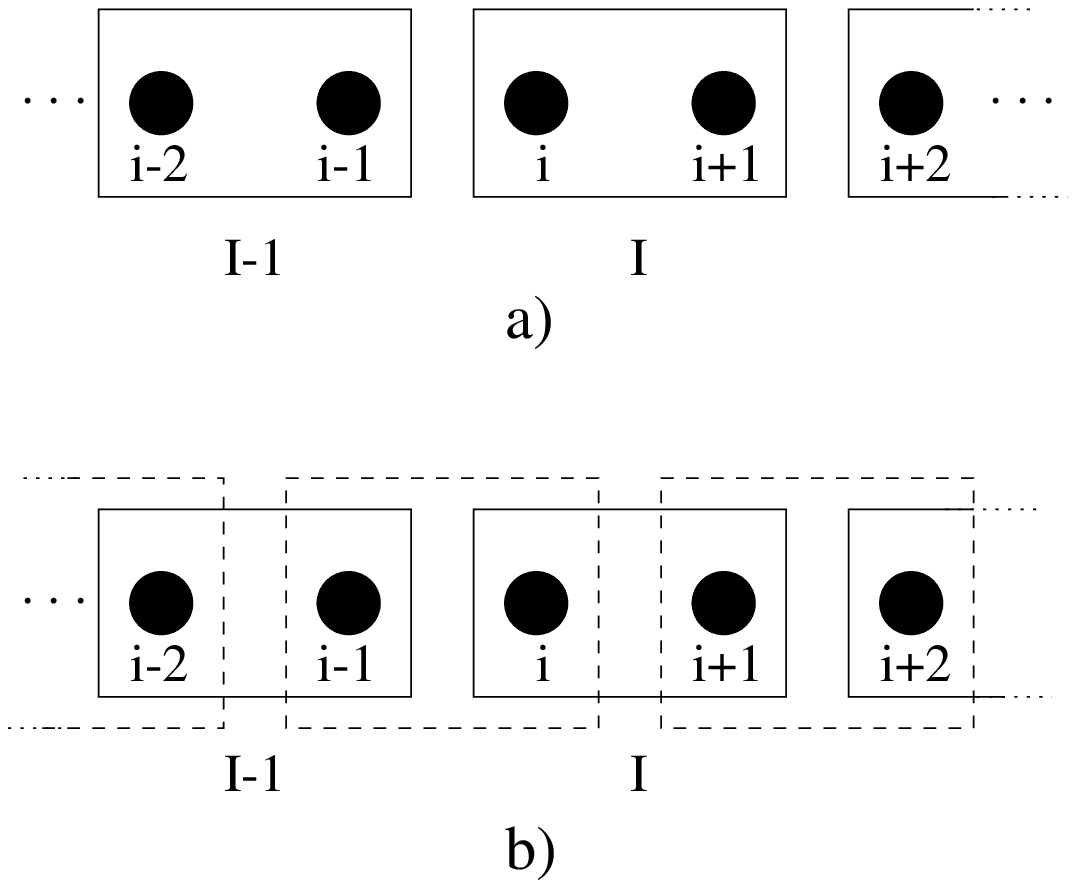,width=9.0cm,height=6.8cm}}
\vspace{0.8cm}
\caption{\em Successive blocks in a one dimensional chain in
   the commuting case a) and the non commuting case b). In the non commuting
   case the system blocks are connected by {\em correlation blocks} drawn by a
   dashed line.}
  \label{correlationblock}
\end{figure}

%
We start this section by giving a very general but well known definition of a
RG transformation (RGT). A RGT $R$ is a map defined on a set of physical
variables $\{{\sigma_l}\}$ and a further set of parameters
${\bf{K}}=(K_1,K_2,\dots)$
\begin{equation}
\label{transf1}
  R:\; \left(\{{\sigma}_l\}, {\bf K}\right)\longrightarrow
     \left(\{{\mu}_m\}, {\bf K}'\right)\;,
\end{equation}
where $\{l\}$ and $\{m\}$ are not necessarily equal indexing sets and
$\{{\mu}_m\}$ denotes the new set of blocked variables belonging to the larger
scale. The quantitative prescription for the map (\ref{transf1}) is then given
in physical terms by including physical constraints as for example the
conservation of symmetries, the maintenance of the structure of the Lagrangian
or the Hamiltonian, or the preservation of physical quantities, like for
example the free energy of the system. Since in most cases it is a difficult
task to define a transformation which combines all needed constraints this
has led to an enormous variety of approximate RG transformations developed in
the last decades \cite{bl82}.\\
The most common realization of the quantitative prescription is to apply
the RG transformation $R$ to the Lagrangian or the Hamiltonian as a
functional which then acts on the variables and parameters given in
(\ref{transf1}). In the special example
of a one dimensional quantum spin chain the new variables are the block spins
and the new coupling constant belongs to the {\em renormalized} set of
parameters ${\bf K}'$. For our case we generalize this RGT to an arbitrary
suitable functional dependence ${\cal O}$
\begin{equation}
\label{transf2}
 R\left[{\cal O}\left(\{{\sigma}_l\}, {\bf K}\right)\right]
  \,=\,{\cal O}\left(\{{\mu}_m\}, {\bf K}'\right)\;.
\end{equation}
By further mathematical analysis of a particular RGT $R$ defined by
(\ref{transf2}) this hopefully yields to a dependence of the {\em renormalized}
parameters ${\bf K}'$ on the old parameters ${\bf K}$ which is called the
{\em flow behaviour} of the RGT. We like to emphasize that once the functional
dependence ${\cal O}\left(\{{\sigma}_l\}, {\bf K}\right)$ is known, we immediately
know the functional dependence ${\cal O}\left(\{{\mu}_m\}, {\bf K}'\right)$
which plays an important role in our construction.\\
We now make the Ansatz that in principle each RGT $R$ can be written as a
composition of two maps, called {\em embedding} and
{\em truncation} \cite{gmdsv95}. This terminology originates from a RG technique
for Hamiltonian systems \cite{jpfd78}, which was then further developed and used
for calculations of the flow behaviour \cite{gmdsv95}.
Rephrasing equation (\ref{transf2}) and focusing only on the renormalization of
the set of parameters for determining the flow behaviour we get
\begin{equation}
\label{transf4}
 G^+\,\circ\,{\cal O}\left({\bf K}\right)\,\circ\, G
  \,=\,{\cal O}\left({\bf K}'\right)
\end{equation}
where we denote $G^+$ as the {\em truncation map} and $G$ as the
{\em embedding map}.\\
As a quite intuitive example for the abstract definition
of the operators $G^+$ and $G$, in the special case of a functional dependence
given by the Hamiltonian, we can construct $G^+$ as a projection map from the
space of all eigenvectors of the Hamiltonian to a space containing a reduced number
of eigenvectors. A projection map from this truncated space back to the space
containing all eigenvectors is a natural way of defining $G$. Although such an
example illustrates a possible application of the abstract formulation given by
(\ref{transf4}) it raises the question of which eigenstates are necessary to keep
for constructing the truncated space. In the case of zero temperature we can
argue that only those eigenvectors should be kept, which correspond to the low
energy eigenvalues \cite{bl82}. As pointed out previously the idea of this paper
should be to invent a real-space RG formulation which overcomes these limitations
by a more abstract formulation.\\
Let us now assume that the functional dependence ${\cal O}$ is given by some
operator, not necessarily the Hamiltonian, on the original Hilbertspace ${\cal H}$
so that equation (\ref{transf4}) can be written as the commuting diagram
\begin{equation}
\label{comm1}
\begin{CD}
{\cal H}' @> G >> {\cal H}\\
@V{\cal O}\left({\bf K}'\right) VV  @VV{\cal O}\left({\bf K}\right) V\\
{\cal H}' @<< G^+ < {\cal H}
\end{CD}
\end{equation}
where ${\cal H}'$ refers to the effective Hilbertspace for the functional
dependence of the truncated set of parameters.\\
We introduce the blocking concept discussed in the previous section as a
tensor product decomposition of the Hilbert space
\begin{equation}
\label{grad1}
{\cal H}\,=\,\underset{I\in {\frak I}}{\bigotimes}\, {\cal H}_{I}\,,
\end{equation}
where ${\frak I}$ denotes some indexing set for the blocks. We are looking for
an embedding and truncation map which respects the block decomposition by
factorization
\begin{align}
\label{grad2}
G_{{\cal H}'}\,=\,\underset{I\in {\frak I}} {\bigotimes}\, G_{{\cal H}'_{I}}
\quad\mbox{and}\quad
G^+_{{\cal H}}\,=\,\underset{I\in {\frak I}} {\bigotimes}\, G^+_{{\cal H}_{I}}\,.
\end{align}
Using this mathematical formulation of the blocking scheme we like to write
the RG transformation for a block in an analogous way
\begin{equation}
\label{grad3}
{\cal O}_{{\cal H}'_{I}}\left({\bf K}'\right)
\,=\, G^+_{{\cal H}_{I}}\,\circ\,{{\cal O}_{{\cal H}_{I}}\left({\bf
      K}\right)}\,
  \circ\,G_{{\cal H}'_{I}}\;,
\end{equation}
due to (\ref{grad2}).
But equation (\ref{grad3}) is not an independent relation since we have to
relate it to the {\em global} relation (\ref{transf4}). To decompose
(\ref{transf4}) into the blocked pieces (\ref{grad3}) we have to assume that
the operator ${\cal O}_{\cal H}$ can be decomposed into commuting block
operators ${\cal O}_{{\cal H}_{I}}$ which is not the case in general in quantum
physics. Therefore the problem encountered so far is to find suitable
functions ${\cal O}_{\cal H}\left({\bf K}\right)$ which respect the block
decomposition of the Hilbert space within the RGT.\\
To find a solution for this problem our Ansatz is to enlarge the Hilbert space
${\cal H}$ by an additional (auxiliary) Hilbert space ${\cal H}_{\text{aux}}$
due to the composition rule
\begin{equation}
\label{total}
 {\cal H}_{\text{total}} \,=\, {\cal H}\,\otimes\,{\cal H}_{\text{aux}}\;.
\end{equation}
We like to think of the space ${\cal H}_{\text{total}}$ as some kind of
'super space' and the global operator
${\cal O}_{{\cal H}\otimes {\cal H}_{\text{aux}}}\left({\bf K}\right)$
is then embedded into the total space ${\cal H}_{\text{total}}$.
The key idea is to recover a block decomposition for
${\cal O}_{{\cal H}\otimes {\cal H}_{\text{aux}}}\left({\bf K}\right)$
into blocked pieces of the form
${{\cal O}_{{\cal H}_{I}\otimes{\left({\cal H}_{\text{aux}}\right)}_{I}}
  \left({\bf K}\right)}$
which we identify as superblocks according to section \ref{dmrgintro}.
The next step in our approach following the basic principles of DMRG is to
outline a general construction for
${\cal O}_{{\cal H}\otimes {\cal H}_{\text{aux}}}\left({\bf K}\right)$
with a commuting block decomposition. This can be performed explicitly by starting
with standard real-space RG concepts.\\
In the formulation of {\em standard} block RG we consider a decomposition of
${\cal O}_{{\cal H}\otimes {\cal H}_{\text{aux}}}\left({\bf K}\right)$
into disconnected block functions given by
\begin{align}
\label{grad4b}
{\cal O}_{{\cal H}_{I}
  \otimes{\left({\cal H}_{\text{aux}}\right)}_{I}}({\bf K})
 \,=\,{\cal O}^{\text{system}}_{{\cal H}_{I}
  \otimes{\left({\cal H}_{\text{aux}}\right)}_{I}}({\bf K})
   \qquad\text{with}\quad I\in {\frak I}\,.
\end{align}
where we have neglected the non commutativity or {\em correlations} between the
blocks completely. A straight forward way to improve the standard RG method
is to include somehow the correlations between adjacent system blocks. As
visualized in figure \ref{correlationblock} we can refer to these correlations as
blocks, which we denote therefore as {\em correlation blocks}.
%
%
Using these correlation blocks enables us to represent the non
commutativities between the system blocks in a compact way and we denote these
correlation blocks, using the overall notation given in the appendix, as
\begin{align}
\label{grad5}
{\cal O}^{\text{correlation}}_{{\cal H}_{\left\{i,i-1,\dots\right\}}
  \otimes{\left({\cal H}_{\text{aux}}\right)}_{\left\{i,i-1,\dots\right\}}}
   ({\bf K})
\end{align}
\begin{align}
  \text{with}\quad {\cal H}_{\left\{i,i-1,\dots\right\}}
   \otimes{\left({\cal H}_{\text{aux}}\right)}_{\left\{i,i-1,\dots\right\}}
  \;\subset\;
   {{\cal H}_{I}\otimes {\left({\cal H}_{\text{aux}}\right)}_{I}
     \otimes {\cal H}_{I-1}\otimes {\left({\cal H}_{\text{aux}}\right)}_{I-1}
      \otimes \dots}\;\;.\nonumber
\end{align}
The subspace
\begin{align}
\label{grad5b}
  {\cal H}_{\left\{i,i-1,\dots\right\}}
   \otimes{\left({\cal H}_{\text{aux}}\right)}_{\left\{i,i-1,\dots\right\}}
  \;=\;
   {\cal H}_{i}\otimes {\left({\cal H}_{\text{aux}}\right)}_{i}
     \otimes {\cal H}_{i-1}\otimes {\left({\cal H}_{\text{aux}}\right)}_{i-1}
      \otimes \dots
\end{align}
denotes the tensor product composition of all the block Hilbert spaces
used for the construction of the correlation block.
%
\section{Decomposition rules}
\label{decom}
%
We are now dealing with the problem how to include these correlation blocks
into the RG transformation. One can find approaches in the past where this
is performed perturbatively \cite{gmdsv95} and therefore unsuitable in our
case. To find some insight into this problem let us start with the composition
\begin{align}
\label{decom1}
{\cal O}_{{\cal H}\otimes{\cal H}_{\text{aux}}}\left({\bf K}\right)
 \,=\,
 \sum_{I\in {\frak I}}{\cal O}^{\text{system}}_{{\cal H}_{I}
  \otimes{\left({\cal H}_{\text{aux}}\right)}_{I}}\left({\bf K}\right)
 \,+\,\sum_{\substack{\left\{i,i-1,\dots\right\}\\[0.1cm] \subset \left\{I,I-1,\dots\right\}}}
  {\cal O}^{\text{correlation}}_{{\cal H}_{\left\{i,i-1,\dots\right\}}
    \otimes{\left({\cal H}_{\text{aux}}\right)}_{\left\{i,i-1,\dots\right\}}}
     \left({\bf K}\right)
\end{align}
which is exact and as always ${\left\{i,i-1,\dots\right\}}$ denotes the subset
of all needed product subspaces for constructing the correlation blocks. We like to
stress that the decomposition (\ref{decom1}) in sums of system blocks and correlation
blocks is not unique. Later we will examine another decomposition to which in contrast
we will refer to as the product decomposition.\\
Let us apply the RG transformation (\ref{transf4}) on the sum decomposition
(\ref{decom1})
\begin{align}
\label{decom2}
{\cal O}_{{\cal H}'\otimes{\cal H}'_{\text{aux}}}\left({\bf K}'\right)
 \,&=\,
 G^+_{{\cal H}\otimes{\cal H}_{\text{aux}}}\,\circ\,\left[
 \sum_{I\in {\frak I}}{\cal O}^{\text{system}}_{{\cal H}_{I}
  \otimes{\left({\cal H}_{\text{aux}}\right)}_{I}}\left({\bf K}\right)\right]
  \,\circ\, G_{{\cal H}'\otimes{\cal H}'_{\text{aux}}}\nonumber\\[0.3cm]
  &\;+\,
 G^+_{{\cal H}\otimes{\cal H}_{\text{aux}}}\,\circ\,\left[
   \sum_{\substack{\left\{i,i-1,\dots\right\}\\[0.1cm] \subset
       \left\{I,I-1,\dots\right\}}}
  \!{\cal O}^{\text{correlation}}_{{\cal H}_{\left\{i,i-1,\dots\right\}}
   \otimes{\left({\cal H}_{\text{aux}}\right)}_{\left\{i,i-1,\dots\right\}}}
    \left({\bf K}\right)
  \right]\,\circ\, G_{{\cal H}'\otimes{\cal H}'_{\text{aux}}}
  \nonumber\\[0.3cm]
\end{align}
and all quantities are used in the context of additional auxiliary spaces.
Let us first consider the system block summand in (\ref{decom2}), which can be
rewritten as\\
\begin{align}
\label{decom3}
\!\!\!\!\!\!\!\!\!\!\!\!\!\!\!\!\!\!\!\!\!\!\!\!\!\!\!
 \sum_{I\in {\frak I}}&\left[ G^+_{{\cal H}\otimes{\cal H}_{\text{aux}}}
  \,\circ\, {\cal O}^{\text{system}}_{{\cal H}_{I}
  \otimes{\left({\cal H}_{\text{aux}}\right)}_{I}}\!\left({\bf K}\right)
  \,\circ\, G_{{\cal H}'\otimes{\cal H}'_{\text{aux}}}\right] =\nonumber\\[0.2cm]
 &\qquad\qquad\sum_{I\in {\frak I}}
  \left[G^+_{{\cal H}_{I}\,\otimes\,
   {\left({\cal H}_{\text{aux}}\right)}_{I}}\,\circ\,
     {{\cal O}^{\text{system}}_{{\cal H}_{I}\otimes
         {\left({\cal H}_{\text{aux}}\right)}_{I}}
      \!\left({\bf K}\right)}\,\circ\, G_{{\cal H}'_{I}\,\otimes\,
        {\left({\cal H}'_{\text{aux}}\right)}_{I}}\right]\nonumber\\[0.2cm]
    &\qquad\qquad\qquad\qquad\cdot\prod_{\substack{J\in {\frak I}
        \\[0.1cm] J\not= I}}
  G^+_{{\cal H}_{J}\,\otimes\, {\left({\cal H}_{\text{aux}}\right)}_{J}}
  \,\circ\,
  G_{{\cal H}'_{J}\,\otimes\, {\left({\cal H}'_{\text{aux}}\right)}_{J}}\;.
\end{align}
This is exactly the local RGT for the system blocks (\ref{grad3}) if we neglect
the last product term on the right hand side of (\ref{decom3}). We will refer
to this factor as a correction term that vanishes if we demand
\begin{align}
\label{decom4}
 G^+_{{\cal H}\otimes{\cal H}_{\text{aux}}}\,\circ\,
   G_{{\cal H}'\otimes{\cal H}'_{\text{aux}}}
   \,=\,{\mathbbm{1}}_{{\cal H}'\otimes{\cal H}'_{\text{aux}}}\;.
\end{align}
Inserting this constraint into (\ref{decom3}), carrying out the same
calculation for the correlation blocks and finally using relation (\ref{grad3})
we get the renormalized version of equation (\ref{decom1}) given by
\begin{align}
\label{decom5}
{\cal O}_{{\cal H}'\otimes{\cal H}'_{\text{aux}}}\left({\bf K}'\right)
 \,=\,
 \sum_{I\in {\frak I}}{\cal O}^{\text{system}}_{{\cal H}'_{I}
  \otimes{\left({\cal H}'_{\text{aux}}\right)}_{I}}\left({\bf K}'\right)
 \,+\sum_{\substack{\left\{i,i-1,\dots\right\}\\[0.1cm] \subset
       \left\{I,I-1,\dots\right\}}}
  \!{\cal O}^{\text{correlation}}_{{\cal H}'_{\left\{i,i-1,\dots\right\}}
    \otimes{\left({\cal H}'_{\text{aux}}\right)}_{\left\{i,i-1,\dots\right\}}}
     \left({\bf K}'\right)\,,
\end{align}
which leads to the renormalized set of parameters ${\bf K}'$. In (\ref{decom5})
we used the reasonable definition
\begin{equation}
\label{grad6}
\begin{split}
{\cal O}^{\text{correlation}}_{{\cal H}'_{\left\{i,i-1,\dots\right\}}
  \otimes{\left({\cal H}'_{\text{aux}}\right)}_{\left\{i,i-1,\dots\right\}}}
  ({\bf K}')
\,:=\,& \left[G^+_{{\cal H}_{\left\{i,i-1,\dots\right\}}\,\otimes\,
    {\left({\cal H}_{\text{aux}}\right)}_{\left\{i,i-1,\dots\right\}}}
\right]\\[0.3cm]
   &\!\!\!\!\!\!\!\!\!\!\!\!\!\!\!\!\!\!\!\!\!\!\!\!\!\!\circ\,
    {{\cal O}^{\text{correlation}}_{{\cal H}_{\left\{i,i-1,\dots\right\}}\otimes
  {\left({\cal H}_{\text{aux}}\right)}_{\left\{i,i-1,\dots\right\}}}
     \!\left({\bf K}\right)}
      \,\circ\,\left[G_{{\cal H}'_{\left\{i,i-1,\dots\right\}}\,\otimes\,
    {\left({\cal H}'_{\text{aux}}\right)}_{\left\{i,i-1,\dots\right\}}}\right]
  \;.\\[0.2cm]
\end{split}
\end{equation}
Relation (\ref{decom4}) introduces an additional constraint for the RGT and
therefore restricts the variety of possible transformations.\\
In the case of a product decomposition of the operator ${\cal O}({\bf K})$ we
can write\\
\begin{align}
\label{2ndcase1}
{\cal O}_{{\cal H}\otimes{\cal H}_{\text{aux}}}\left({\bf K}\right)
 \,=\,
 \prod_{i\in I}{\cal O}^{\text{system}}_{{\cal H}_{I}
  \otimes{\left({\cal H}_{\text{aux}}\right)}_{I}}\left({\bf K}\right)
 \;\,\cdot\!\!\prod_{\substack{\left\{i,i-1,\dots\right\}\\[0.1cm] \subset
       \left\{I,I-1,\dots\right\}}}
  \!{\cal O}^{\text{correlation}}_{{\cal H}_{\left\{i,j,\dots\right\}}
    \otimes{\left({\cal H}_{\text{aux}}\right)}_{\left\{i,j,\dots\right\}}}
     \left({\bf K}\right)\,.
\end{align}
In analogy to the case of the sum decomposition (\ref{decom1}) we can apply
the RG transformation (\ref{transf4}) to (\ref{2ndcase1}) which leads to the
expression
\begin{align}
\label{2ndcase2}
&{\cal O}_{{\cal H}'\otimes{\cal H}'_{\text{aux}}}\!\!\left({\bf K}'\right)
 \nonumber\\[-0.1cm] & \quad =
 G^+_{{\cal H}\otimes{\cal H}_{\text{aux}}}\circ\left[
 \prod_{i\in I}{\cal O}^{\text{system}}_{{\cal H}_{I}
  \otimes{\left({\cal H}_{\text{aux}}\right)}_{I}}\left({\bf K}\right)
 \cdot\!\!\!\!\prod_{\substack{\left\{i,i-1,\dots\right\}\\[0.1cm] \subset
       \left\{I,I-1,\dots\right\}}}
  \!\!\!\!\!\!{\cal O}^{\text{correlation}}_{{\cal H}_{\left\{i,j,\dots\right\}}
    \otimes{\left({\cal H}_{\text{aux}}\right)}_{\left\{i,j,\dots\right\}}}
     \left({\bf K}\right)\right]
  \circ G_{{\cal H}'\otimes{\cal H}'_{\text{aux}}}\;.
\end{align}
Since this is already the final step in the calculation for this special case
of a decomposition we are not able to write the result as a composition of the
renormalized system block part and correlation block part as we did in
(\ref{decom5}) for the sum decomposition. By the considerations so far the 
product decomposition therefore seems to be not as useful as the sum decomposition
for later applications. This is not the case as we will show in the following.\\
For the auxiliary space we distinguish between two different cases, an
{\em active} role and a {\em passive} role. Here active means that the
auxiliary space is directly involved into the RGT, i.e. $G$ and $G^+$ act
nontrivial on this additional space. The commutative diagram describing the
general active situation is given in (\ref{comm2}).\\
\begin{equation}
\label{comm2}
\begin{CD}
{\cal H}'\otimes{\cal H}'_{\text{aux}}
 @> G_{{\cal H}'\otimes{\cal H}'_{\text{aux}}} >> {\cal H}\otimes{\cal H}_{\text{aux}}\\
@V{{\cal O}_{{\cal H}'\otimes{\cal H}'_{\text{aux}}}\left({\bf K}'\right)}
 VV  @VV{{\cal O}_{{\cal H}\otimes{\cal H}_{\text{aux}}}\left({\bf K}\right)}V\\
{\cal H}'\otimes{\cal H}'_{\text{aux}}
 @<<G^+_{{\cal H}\otimes{\cal H}_{\text{aux}}} < {\cal H}\otimes{\cal H}_{\text{aux}}
\end{CD}
\end{equation}
Relation (\ref{comm2}) reduces to a rewriting of (\ref{comm1}), if the
transformation maps $G$ and $G^+$ each operate as the identity on the
auxiliary space and the functional dependence ${\cal O}\left({\bf K}\right)$
acts non trivial only on ${\cal H}$. This gives us an example of the particular
case of a passive role of the auxiliary space as it is depicted in (\ref{comm3}).\\
\begin{equation}
\label{comm3}
\begin{CD}
{\cal H}'\otimes{\cal H}'_{\text{aux}}
 @> G_{{\cal H}'}\otimes{\mathbbm{1}}_{{\cal H}'_{\text{aux}}} >> {\cal H}\otimes{\cal H}_{\text{aux}}\\
@V{{\cal O}_{{\cal H}'\otimes{\cal H}'_{\text{aux}}}\left({\bf K}'\right)}
 VV  @VV{{\cal O}_{{\cal H}\otimes{\cal H}_{\text{aux}}}\left({\bf K}\right)}V\\
{\cal H}'\otimes{\cal H}_{\text{aux}}
 @<<G^+_{\cal H}\otimes{\mathbbm{1}}_{{\cal H}_{\text{aux}}} < {\cal H}\otimes{\cal H}_{\text{aux}}
\end{CD}
\end{equation}
In the case of (\ref{comm2}) we can think of the auxiliary space as some kind
of medium not changed during a RG step. The active and the passive choice of the
auxiliary space yield two different realizations of our RG, which we will call
the 'general (real-space) RG' (GRG) and refer to the corresponding RG transformation
as GRGT.
%
\section{The construction of the local GRG transformation}
\label{constr}
%
So far we have discussed different types of quantum decompositions and types of
auxiliary spaces. We now turn to the question how to construct the embedding map
$G_{{\cal H}'\otimes{\cal H}'_{\text{aux}}}$ and the truncation map
$G^+_{{\cal H}\otimes{\cal H}_{\text{aux}}}$. In (\ref{transf2}) we used the
functional dependence ${\cal O}$ to introduce physical constraints within the
RG transformation. To determine $G_{{\cal H}'\otimes{\cal H}'_{\text{aux}}}$
and $G^+_{{\cal H}\otimes{\cal H}_{\text{aux}}}$ we introduce another
constraint. In addition to keeping the structure of the operator ${\cal O}$ we
relate ${\cal O}$ to a physical quantity $\cal Z(\cal O)$ which acts as a
physical invariant $\!\!$\renewcommand{\thefootnote}{\dag}\footnote{A
  possible example for such a quantity can be the
  partition function or the free energy of the physical system.}.
Equating the original physical quantity $\cal Z(\cal O)$ calculated from the
original quantum lattice and the effective physical quantity $\cal Z(\cal O')$
for the reduced lattice we obtain
$G^+_{{\cal H}\otimes{\cal H}_{\text{aux}}}$ and
$G_{{\cal H}'\otimes{\cal H}'_{\text{aux}}}$ from
\begin{equation}
\label{inveq}
{\cal Z}\left[{\cal O}_{{\cal H}\otimes{\cal H}_{\text{aux}}}
  \left({\bf K}\right)\right]\,=\,{\cal Z}\left[
  G^+_{{\cal H}\otimes{\cal H}_{\text{aux}}}\,\circ\,
  {\cal O}_{{\cal H}\otimes{\cal H}_{\text{aux}}}\left({\bf K}\right)\,\circ\,
  G_{{\cal H}\otimes{\cal H}_{\text{aux}}}\right]
\,=\,{\cal Z}\left[{\cal O}_{{\cal H}'\otimes{\cal H}'_{\text{aux}}}
  \left({\bf K}'\right)\right]\;.
\end{equation}
We refer to equation (\ref{inveq}) as the {\em {invariance relation}} for the
RGT. Finally we have to decompose $G^+_{{\cal H}\otimes{\cal H}_{\text{aux}}}$
and $G_{{\cal H}'\otimes{\cal H}'_{\text{aux}}}$ according to (\ref{grad2}).\\
We are now able to give the precise definition of the local RGT in the form
\begin{equation}
\label{comm4}
\begin{CD}
  {\cal H}'_{I}\otimes {\left({\cal H}'_{\text{aux}}\right)}_{I} @> G_{{\cal
      H}'_{I}\,\otimes\,{\left({\cal H}'_{\text{aux}}\right)}_{I}} >>
  {\cal H}_{I}\otimes {\left({\cal H}_{\text{aux}}\right)}_{I}\\ @V {{\cal
      O}_{{\cal H}'_{I}\otimes{\left({\cal H}'_{\text{aux}}\right)}_{I}}
    \left({\bf K}'\right)} VV @VV {{\cal O}_{{\cal H}_{I}\otimes{\left({\cal
            H}_{\text{aux}}\right)}_{I}} \left({\bf K}\right)} V \\ {\cal
      H}'_{I}\otimes {\left({\cal H}'_{\text{aux}}\right)}_{I} @<< G^+_{{\cal
        H}_{I}\,\otimes\,{\left({\cal H}_{\text{aux}}\right)}_{I}} <
    {\cal H}_{I}\otimes {\left({\cal H}_{\text{aux}}\right)}_{I}\\
\end{CD}
\end{equation}
where we refer to
$G^+_{{\cal H}_{I}\,\otimes\,{\left({\cal H}_{\text{aux}}\right)}_{I}}$ and
$G_{{\cal H}'_{I}\,\otimes\,{\left({\cal H}'_{\text{aux}}\right)}_{I}}$ as the
generators of the transformation. By the explanations of section \ref{decom}
\begin{align}
\label{prod1l}
{\cal O}_{{\cal H}_{I}
  \otimes{\left({\cal H}_{\text{aux}}\right)}_{I}}\left({\bf K}\right)
 \,=\,
 {\cal O}^{\text{system}}_{{\cal H}_{I}
  \otimes{\left({\cal H}_{\text{aux}}\right)}_{I}}\left({\bf K}\right)
 \;\,\cdot\!\!\prod_{\substack{\left\{i,i-1,\dots\right\}\\[0.1cm] \subset I}}
  \!{\cal O}^{\text{correlation}}_{{\cal H}_{\left\{i,j,\dots\right\}}
    \otimes{\left({\cal H}_{\text{aux}}\right)}_{\left\{i,j,\dots\right\}}}
     \left({\bf K}\right)
\end{align}
\begin{align}
\label{prod2l}
\text{or}\qquad{\cal O}_{{\cal H}_{I}
  \otimes{\left({\cal H}_{\text{aux}}\right)}_{I}}\left({\bf K}\right)
 \,=\,
 {\cal O}^{\text{system}}_{{\cal H}_{I}
  \otimes{\left({\cal H}_{\text{aux}}\right)}_{I}}\left({\bf K}\right)
 \;\,+\!\!\sum_{\substack{\left\{i,i-1,\dots\right\}\\[0.1cm] \subset I}}
  \!{\cal O}^{\text{correlation}}_{{\cal H}_{\left\{i,j,\dots\right\}}
    \otimes{\left({\cal H}_{\text{aux}}\right)}_{\left\{i,j,\dots\right\}}}
  \left({\bf K}\right)\,
\end{align}
and analogously for
${{\cal O}_{{\cal H}'_{I}\otimes{\left({\cal H}'_{\text{aux}}\right)}_{I}}
    \left({\bf K}'\right)}$.
%
\section{Perfect and exact local RG transformations}
\begin{figure}[ht]
\vspace{0.6cm}
\centerline{
\psfig{figure=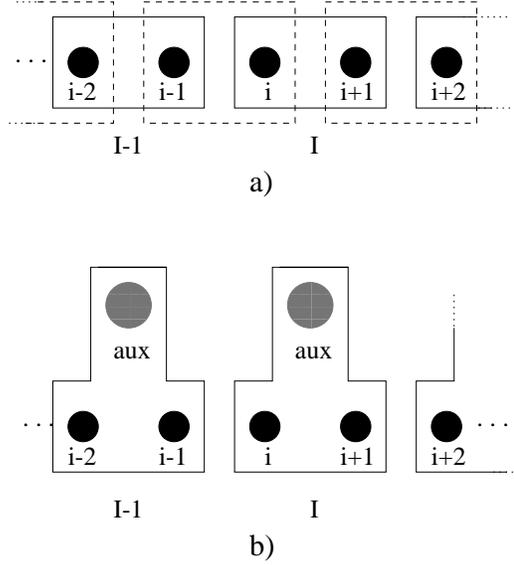,width=7.0cm,height=7.4cm}}
\vspace{0.8cm}
\caption{\em A rigorous blocked chain in the non commuting case with a)
  decomposition into system and correlation blocks, the latter visualized by dashed
 boxes and b) decomposition into system blocks, each equipped with an auxiliary
 space suitable to take over the role of the correlation blocks during the RGT.}
  \label{auxpic}
\end{figure}

%
In this section we study the relationship between (\ref{comm4}) and the
global RGT\\
\begin{equation}
\label{glob1}
\begin{CD}
  {\cal H}'\otimes {{\cal H}'_{\text{aux}}} @> G_{{\cal
      H}'\,\otimes\,{{\cal H}'_{\text{aux}}}} >>
  {\cal H}\otimes {{\cal H}_{\text{aux}}}\\ @V {{\cal
      O}_{{\cal H}'\otimes{{\cal H}'_{\text{aux}}}}
    \left({\bf K}'\right)} VV @VV {{\cal O}_{{\cal H}\otimes{\left({\cal
            H}_{\text{aux}}\right)}} \left({\bf K}\right)} V \\ {\cal
      H}'\otimes {{\cal H}'_{\text{aux}}} @<< G^+_{{\cal
        H}\,\otimes\,{{\cal H}_{\text{aux}}}} <
    {\cal H}\otimes {{\cal H}_{\text{aux}}}\\
\end{CD}
\end{equation}
Diagram (\ref{glob1}) represents an exact relation which implies all the
necessary constraints for the RG procedure as can be verified from equation
(\ref{inveq}). We therefore choose relation (\ref{glob1}) as the basic relation
in defining local RGTs.\\
Decomposing the global RGT (\ref{glob1}) into local RGTs given by
(\ref{comm4}) demands for a decomposition of $\cal O$ into commuting
blocks. From previous considerations we conclude that this is impossible
for quantum chains due to the correlation blocks occurring in a decomposition
of a quantum physical system. Therefore the idea is to use the auxiliary
space to decompose the chain into commuting blocks by storing the information
about the correlations of adjacent system blocks within the auxiliary space. By
the decompositions discussed so far we then decompose a chain into system blocks
and try to find an auxiliary space ${\left({\cal H}_{\text{aux}}\right)}_{I}$
for each system block which takes over the role of the correlation blocks within the
RGT as visualized in figure \ref{auxpic}.
%
%
This statement can be made more precise.
\begin{defn}
A local RGT is said to be {\em perfect} if there exists a local operator\\
\begin{align*}
{\cal O}_{{\cal H}'_{I}\otimes
   {\left({\cal H}'_{\text{aux}}\right)}_{I}}
 \left({\bf K}'\right) \,=\, \left[G^+_{{\cal H}_{I}\,\otimes\,
   {\left({\cal H}_{\text{aux}}\right)}_{I}}\right]\,\circ\,
   {{\cal O}^{\text{system}}_{{\cal H}_{I}\otimes
   {\left({\cal H}_{\text{aux}}\right)}_{I}} \!\left({\bf K}\right)}\,\circ\,
    \left[G_{{\cal H}'_{I}\,\otimes\,
     {\left({\cal H}'_{\text{aux}}\right)}_{I}}\right]
\end{align*}
together with a global functional dependence\,
${\cal \tilde O}_{{\cal H}'\otimes{\cal H}'_{\text{aux}}}\left({\bf K}'\right)$\,
defined by the decomposition\\
\begin{align*}
{\cal \tilde O}_{{\cal H}'\otimes{\cal H}'_{\text{aux}}}\left({\bf K}'\right)
 \,:=\,
 \sum_{I\in {\frak I}}{\cal O}_{{\cal H}'_{I}
  \otimes{\left({\cal H}'_{\text{aux}}\right)}_{I}}\left({\bf K}'\right)
\quad\;{\text {or}}\quad\;
{\cal \tilde O}_{{\cal H}'\otimes{\cal H}'_{\text{aux}}}\left({\bf K}'\right)
 \,:=\,
 \prod_{I\in {\frak I}}{\cal O}_{{\cal H}'_{I}
  \otimes{\left({\cal H}'_{\text{aux}}\right)}_{I}}\left({\bf K}'\right)
\end{align*}
and {\bf no} further local relation governing the renormalization of the correlation
block part occurs.
\end{defn}
The main advantage of a perfect RGT is a rigorous mathematical
description for a local RGT. Although the structure of the local Operator
${\cal O}_{{\cal H}_{I}\otimes{\left({\cal H}_{\text{aux}}\right)}_{I}}$
is conserved, a perfect RGT does not make use of the invariance relation
(\ref{inveq}).
\begin{defn}
A local RGT is said to be {\em exact} if it is perfect and\\
\begin{align*}
{\cal Z}\left[{\cal O}_{{\cal H}\otimes{\cal H}_{\text{aux}}}
  \left({\bf K}\right)\right]
 \,=\,{\cal Z}\left[{\cal O}_{{\cal H}'\otimes{\cal H}'_{\text{aux}}}
  \left({\bf K}'\right)\right]
 \,=\,{\cal Z}\left[{\cal \tilde O}_{{\cal H}'\otimes{\cal H}'_{\text{aux}}}
  \left({\bf K}'\right)\right]\;.
\end{align*}
\end{defn}
If a RGT is exact it includes all needed
constraints and therefore we can compare the RGT to the classical situation
where non commutativity effects are absent.\\
At this point we like to give some important remarks on perfect and exact RGTs.
Although in both cases a rigorous mathematical formalism is used, a physical
approximation usually enters the problem by choosing an appropriate auxiliary
space. Only for a certain class of models we will be able to find auxiliary
spaces with a structure that allows for describing the non commutativity effects
without any approximation.\\
We stress again that in the exact as well as in the perfect RGT
${\cal O}_{\cal H}$ and ${\cal O}_{{\cal H}'}$ are known so that we can
determine $G$ and $G^+$ in both cases according to the explanations in section
\ref{constr}.\\
If the auxiliary space is active it may happen that it vanishes by truncation
during the RG procedure. In such a case no auxiliary space is available after
the local transformation has been worked out and the previously provided information
concerning the correlations between adjacent system blocks is lost.
Therefore the RGT is at most perfect.\\
In the case of an auxiliary space which (only) allows for an approximate
description of the correlations between adjacent system blocks we would like to
have some insight into the accuracy of the approximation. Here we remember the
numerical DMRG procedure in which convergence of numerical values of
ground state quantities by enlarging the superblock is used as an estimate for
the accuracy of the method.\\
It is apparent that only in the case of an exact RGT we are able to calculate
global quantities like the total ground state energy shift. Since we are mainly
interested in an overall effective coupling determining the RG flow we are
looking for exact RGTs.
%
\section{Conclusions}
%
We have invented a non perturbative quantum RG method based on the idea of an
additional auxiliary space. The work was motivated by the success of the DMRG
concerning numerical results and the open question of an underlying general
mathematical framework.\\
The main objects introduced in this article are the auxiliary space
${\cal H}_{\text{aux}}$ and the two maps
$G^+_{{\cal H}_{I}\,\otimes\,{\left({\cal H}_{\text{aux}}\right)}_{I}}$ and
$G_{{\cal H}'_{I}\,\otimes\,{\left({\cal H}'_{\text{aux}}\right)}_{I}}$
which generate the RGT. By using these quantities we were able to give the
definition of an exact local RGT which is the final result of this work. An
exact local RGT involves all the information provided by the physical system.\\
In future work we will proceed by applying our abstract formalism presented
here to quantum spin chains like the Heisenberg models and compare our results
in the context of related work on these models \cite{so96}. This leads us to
concrete and different examples of possible auxiliary spaces. As expected, the
correct choice of the auxiliary space will be the main ingredient in the
construction of the RGT, whereas the definition of the maps
$G^+_{{\cal H}_{I}\,\otimes\,{\left({\cal H}_{\text{aux}}\right)}_{I}}$ and
$G_{{\cal H}'_{I}\,\otimes\,{\left({\cal H}'_{\text{aux}}\right)}_{I}}$ turns
out to be rather straight forward. We also hope for further applications of
the method introduced here.\\[0.8cm]
%
{\leftline{\large\bf Acknowledgment}}\\[0.2cm]
I'm grateful to my colleagues and friends Javier Rodriguez Laguna, Johannes
G\"ottker-Schnetmann and Juri Rolf for encouraging me to proceed with this
work. I would like to thank Prof. P. Stichel for helpful discussions and for
reading the manuscript.\\[1.0cm]
%
\centerline{\large\bf Appendix}\\[0.4cm]
%
Throughout this work blocks are denoted by capital indexing letters,
corresponding to the block sites. The indexing set for the blocks is denoted
as ${\frak I}$. Neighbouring blocks are denoted by a sequence
$I, I-1, I-2, \dots \in {\frak I}$ whereas arbitrary blocks are indexed by
different letters $I, J, \dots \in {\frak I}$.\\
A block Hilbert space ${\cal H}_I$ contains at minimum two single site
Hilbert spaces ${\cal H}_i$ and ${\cal H}_{i-1}$. Single site
Hilbert spaces are denoted by letters $i,j,k,\dots$. To point out that
a single site space ${\cal H}_i$ is contained in a block space ${\cal H}_I$ we
write ${\cal H}_i \subset{\cal H}_I$ or even simpler $i\in I$ if it is clear
that $I$ refers to the block Hilbert space. We also use the abbreviation
$\{i,i-1,\dots\}\subset\{I,I-1,\dots\}$ instead of writing
${\cal H}_{\{i,i-1,\dots\}}\subset {\cal H}_{\{I\}}
\otimes{\cal H}_{\{I-1\}},\dots$. By this notation it becomes not clear which
single site space is contained in a certain block Hilbert space. If
this is important it must be pointed out explicitly.\\[0.1cm]
Expressions which are written in the form {\em expression} are either defined
and used in this work or have special physical meaning.\\[0.1cm]
%
%

\end{document}